\documentclass[cits]{PoS}
\usepackage{amsmath}
\usepackage{amsfonts}
\usepackage{amssymb}
\usepackage{graphicx}
\usepackage{longtable}

\DeclareMathOperator{\tr}{tr}

\newcommand{\Sym}{\textrm{S}}
\newcommand{\AS}{\textrm{AS}}
\newcommand{\Adj}{\textrm{Adj}}

\title{Orientifold Planar Equivalence: The Quenched Meson Spectrum}

\ShortTitle{Orientifold Planar Equivalence: The Quenched Meson Spectrum}

\author{\speaker{Biagio Lucini}\\
        School of Physical Sciences, Swansea University,
        Singleton Park, Swansea SA2 8PP, UK\\
        E-mail: \email{b.lucini@swansea.ac.uk}}

\author{Gregory Moraitis\\
        School of Physical Sciences, Swansea University,
        Singleton Park, Swansea SA2 8PP, UK\\
        E-mail: \email{pygm@swansea.ac.uk}}

\author{Agostino Patella\\
        CERN, Physics Department, 1211 Geneva 23, Switzerland\\
        E-mail: \email{agostino.patella@cern.ch}}

\author{Antonio Rago\\
        Department of Physics, Bergische Universit\"at Wuppertal, Gaussstr. 20, D-42119 Wuppertal, Germany\\
        E-mail: \email{rago@physik.uni-wuppertal.de}}

\abstract{A numerical study of Orientifold Planar Equivalence is
  performed in SU($N$) Yang-Mills theories for $N=2,3,4,6$. Quenched
  meson masses are extracted in the antisymmetric,  symmetric and
  adjoint representations for the pseudoscalar and vector channels.
  An extrapolation of the vector mass as a function of the
  pseudoscalar mass to the large-$N$ limit shows that the numerical
  results agree within errors for the three theories, as predicted by
  Orientifold Planar Equivalence. As a byproduct of the
  extrapolation, the size of the corrections up to $O(1/N^3)$ are
  evaluated. A crucial prerequisite for the extrapolation is the
  determination of an analytical relationship between the corrections
  in the symmetric and in the antisymmetric representations, order by
  order in a $1/N$ expansion.\\
 [0.5cm]
\rightline{CERN-PH-TH/2010-254}
\rightline{WUB/10-30\hspace{1.8cm}}
 }

\FullConference{The XXVIII International Symposium on Lattice Field Theory, Lattice2010\\
		June 14-19, 2010\\
		Villasimius, Italy}

\begin{document}

\section{Introduction}
Orientifold Planar Equivalence~\cite{Armoni:2003gp} is a powerful analytical
tool that establishes the equality of certain observables (among which, meson
masses) in two classes of gauge theories at $N = \infty$, $N$ being the number
of colours. This equivalence is of particular interest when the two gauge
theory families it relates are SU($N$) with $N_f$ (anti)symmetric Dirac
fermion flavours and SU($N$) with $N_f$ Majorana flavours in the adjoint
representation. In fact, this special case enables us to
transcribe SUSY results to QCD~\cite{Armoni:2003fb}, provided that the latter
is close (in the sense of the $1/N$ expansion) to its large-$N$ limit. Some
issues related to the equivalence, like e.g. the size of the finite $N$
corrections, mandate an {\em ab-initio}
calculation. A full dynamical calculation is expensive from a
computational point of view. In fact, the cost of the computation is
brought almost entirely by the inversion of the fermionic matrix,
which for theories with two-index representation fermions requires $O(N^4)$
operations. Before undertaking a full dynamical study, it is
convenient to consider the quenched case as a prototype
example. Although for theories in two-index representations the
quenched theory and the dynamical theory do not coincide in the
large-$N$ limit, Orientifold Planar Equivalence still holds in the
quenched case, which then becomes a useful toy model to study techniques that
could be used in the dynamical investigation. In this spirit, a first paper
appeared in which the chiral condensate was shown to coincide in
the large-$N$ limit of the adjoint, symmetric and antisymmetric
representations~\cite{Armoni:2008nq}. A crucial point in that work was the
separation of the corrections in even and odd powers of $1/N$ by
combining the numerical results in the symmetric and antisymmetric
representations. As a result, a precise extrapolation to the large-$N$
limit could be performed and Orientifold Planar Equivalence was
verified to a high degree of accuracy. Moreover, this technique
allowed a computation of corrections up to $O(1/N^3)$ in the
(anti)symmetric representation while still using a reasonably small
value of $N$. In this work, we will perform a similar study for the
quenched meson spectrum. First, we will show that there is a simple order
by order relationship between the coefficients in a large-$N$
expansion of amplitudes in correlation functions and of spectral
masses in the symmetric and antisymmetric representations. Then, we
shall use a simple chiral ansatz to establish a relationship between
the mass of the vector meson $m_V$ and the mass of the pseudoscalar
meson $m_{PS}$ at finite $N$, for $N$ ranging from two to six, in the
adjoint, in the symmetric and antisymmetric representations. Finally,
using the analytical relationships between the corrections, the chiral
ansatz will be extrapolated to the large-$N$ limit and Orientifold
Planar Equivalence proven to hold. The results we will present in this
work have already appeared in~\cite{Lucini:2010kj}, to which we refer
for a more detailed discussion.

\section{Meson masses for two-index representation fermions in the $1/N$ expansion}
We extract meson masses from the large distance exponential decay of
the meson correlation functions. For the fermion sources we use the
Wilson formulation, in terms of which the Dirac operator is given by
\begin{eqnarray}
D_{x y}  = \phantom{-} (m+4r) \delta_{xy} - K_{x y} \ ,
\end{eqnarray}
with $m$ the bare quark mass and
\begin{eqnarray}
K_{x y}  = - \frac{1}{2} \left[\left(r -
  \gamma_{\mu}\right) R\left[U_{\mu}^{\phantom{\dag}}(x)\right] \delta_{y,x+\hat{\mu}} +
  \left(r +
  \gamma_{\mu}\right) R\left[U_{\mu}^{\dag}(y)\right]
  \delta_{y,x-\hat{\mu}}  \right]  \ .
\end{eqnarray}
In the previous formulae, $x$ and $y$ are site indices and the
$\gamma_{\mu}$ the Dirac $\Gamma$ matrices in Euclidean space.
$R\left[U_{\mu}(x)\right]$ is the link variable stemming from $x$
in the direction $\hat{\mu}$ expressed in the representation $R$. In our
case, $R$ can be the adjoint ($\Adj$), the two-index symmetric
($\Sym$) or the two-index antisymmetric ($\AS$) representation. For
the representation $R$, meson correlators take the form
\begin{eqnarray}
C_{\Gamma_1 \Gamma_2}^R(x,y) = 
r_{R} \left\langle \tr_R \left(D^{-1}_{y x}
    \Gamma_1^\dag
D^{-1}_{x y} \Gamma_2 \right) \right\rangle_{YM} \ ,
\end{eqnarray}
where $\tr_R$ indicates the trace over the color indices in the
representation $R$ and $\Gamma_1$ and $\Gamma_2$ are combinations of
$\Gamma$ matrices associated with the quantum numbers $J^{PC}$ of the
mesons. With the choice $r_R =1$ for $R=\Sym/\AS$ and $r_R = 1/2$ for $R=\Adj$,
in the large-$N$ limit we have the equality of correlators in these three
representations. The factor $r_R = 1/2$ in the adjoint representation can
be understood considering that in our formulation we are always taking Dirac
flavours in the adjoint, and one Dirac flavour is equivalent to two Majorana flavours. 

The formal proof of the equality of the correlators can be sketched as follows. We start expanding the correlators in Wilson loops:
\begin{flalign}
\label{eq:expansion}
\frac{1}{N^2} C_{\Gamma_1 \Gamma_2}^{R}(x,y) = \frac{r_R}{N^2} \sum_{{\cal C} \supset (x,y)} \alpha_{\cal C} \, \langle \, \tr_R W_{\cal C} \, \rangle
\ ,
\end{flalign}
${\cal C}$ being a closed curve that contains both $x$ and $y$. The
coefficient $\alpha_{\cal C}$ in \eqref{eq:expansion} does not depend on the
representation. Writing explicitly the traces in two-index
representations in terms of the trace of the fundamental
representation yields 
\begin{flalign}
& \frac{1}{N^2} C_{\Gamma_1 \Gamma_2}^{\Sym/\AS}(x,y) = 
\frac{1}{2}  \sum_{{\cal C} \supset (x,y)} \alpha_{\cal C}
\frac{\langle [\tr W_{\cal C}]^2 \rangle \pm \langle \tr [W_{\cal C}^2] \rangle}{ N^2 } \ , \\
& \frac{1}{N^2} C_{\Gamma_1 \Gamma_2}^{\Adj}(x,y) = 
\frac{1}{2}\sum_{{\cal C} \supset (x,y)} \alpha_{\cal C} 
\frac{ \langle |\tr W_{\cal C}|^2 \rangle - 1}{ N^2 } \ .
\end{flalign}
Note that the factors of $1/2$ in the two previous formulae have a different origin: for the $\Sym/\AS$, it comes when
expressing the trace in the two-index representations in terms of the group element in the fundamental representation, while for
the adjoint is the factor $r_{R}$. At large $N$, we can use factorisation and neglect the subleading terms, obtaining
\begin{flalign}
& \frac{1}{N^2} C_{\Gamma_1 \Gamma_2}^{\Sym/\AS}(x,y) = 
\frac{1}{2}  \sum_{{\cal C} \supset (x,y)} \alpha_{\cal C}
\frac{\langle \tr W_{\cal C} \rangle \langle \tr W_{\cal C} \rangle}{ N^2 } \ ,\\
& \frac{1}{N^2} C_{\Gamma_1 \Gamma_2}^{\Adj}(x,y) = 
\frac{1}{2}\sum_{{\cal C} \supset (x,y)} \alpha_{\cal C}
\frac{\langle \tr W_{\cal C}\rangle \langle \tr W_{\cal C}^{\dag} \rangle}{ N^2 } \ .
\end{flalign}
Finally, the equality
\begin{eqnarray}
\lim_{N \rightarrow \infty} \frac{1}{N^2} C_{\Gamma_1 \Gamma_2}^{\Sym/AS}(x,y)
\ = \ 
\lim_{N \rightarrow \infty} \frac{1}{N^2} C_{\Gamma_1 \Gamma_2}^{\Adj}(x,y)
\end{eqnarray}
follows from charge conjugation invariance, which dictates $\langle \tr
W_{\cal C}^{\dag} \rangle  = \langle \tr W_{\cal C} \rangle$. This
last step in Yang-Mills theories does not present the conceptual
issues discussed in~\cite{Unsal:2006pj,Armoni:2007rf} for the dynamical case.
The equality of the meson spectra in the three theories follows from
the equality of the correlators.

Correlators of mesons in the adjoint representation are expected to be
expandable in a power series of $1/N^2$. As for the other two
representations, since in the symmetric and antisymmetric
representations the static sources introduce $1/N$ effects, there is
no reason not to expect $1/N$ corrections in the meson mass spectrum.
Assuming that there is no accidental
degeneracy at large-$N$ in the theory  with fermions in the antisymmetric
and in the symmetric\footnote{In Quantum Field Theory, degeneracies are due
to a symmetry, and there is no symmetry that could generate degeneracies in
these theories.}, the masses $m_{R}$ and the amplitudes $A_{R}$ in correlators
of fermion bilinears in these two representations can be shown to be related by 
\begin{eqnarray}
A^j_{\Sym}(N) = A^j_{\AS}(-N) \qquad \mbox{and} \qquad m^j_{\Sym}(N) = m^j_{\AS}(-N)
\end{eqnarray}
for states $j$ related by Orientifold Planar Equivalence. Order by order in a $1/N$ expansion, this implies that even-power terms in $1/N$ in the symmetric and antisymmetric representation have the same coefficient, while odd-power terms have opposite coefficients. This means that the combinations
\begin{eqnarray}
\label{eq:auxmass}
M^j =  \left( m^j_{\Sym} + m^j_{\AS} \right)/2 \qquad  \mbox{and} \qquad \mu^j = N \left( m^j_{\Sym} - m^j_{\AS} \right)/2
\end{eqnarray}
have an expansion in powers of $1/N^2$. 

\section{Numerical results}
We generated gauge configurations for values of $N$ ranging from two to six using the standard Wilson action. Both the generation
of the configurations and the analysis of the correlators in two-index representations were performed using the purposely
developed HiRep code~\cite{DelDebbio:2008zf}. The technical details of the simulations have been given in~\cite{Lucini:2010kj}. Here
we remark that we simulate at the same $\beta$ values as in
Ref.~\cite{Armoni:2008nq}, where the couplings were chosen so that the gauge
theories at various $N$ are at a common value of the lattice spacing
$a = 1/(5 T_c)$, $T_c$ being the deconfinement temperature of the pure
gauge system ~\cite{Lucini:2003zr}.

\begin{figure}
\begin{center}
\includegraphics*[width=.6\textwidth]{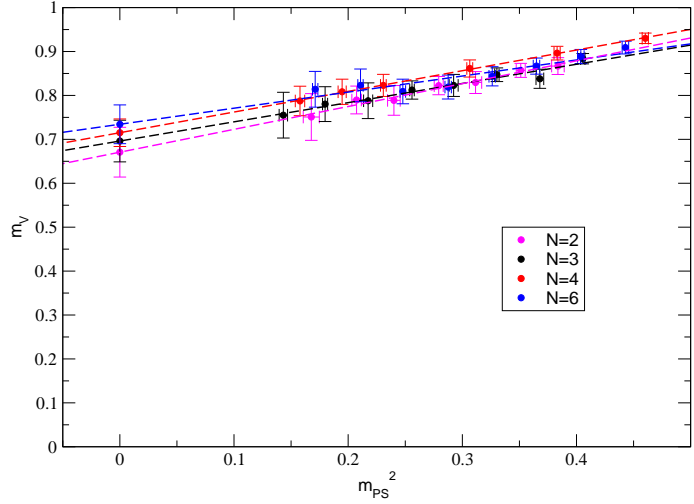}
\caption{$m_V$ as a function of $m_{PS}^2$ for the adjoint
  representation at the values of $N$ shown. The solid lines are
  chiral fits to the data.}
\label{fig:1}
\end{center}
\end{figure}

\begin{figure}
\begin{center}
\includegraphics*[width=.6\textwidth]{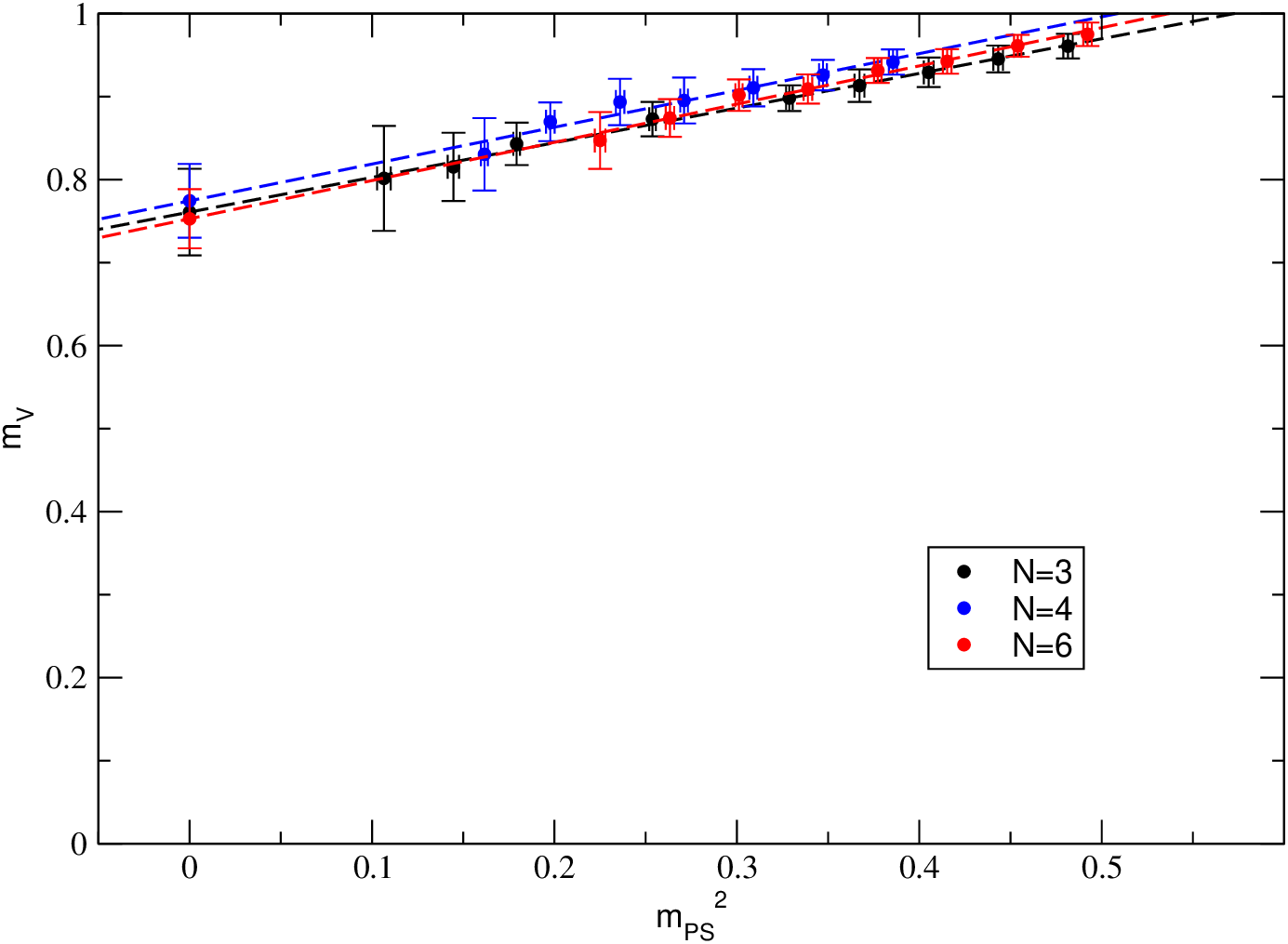}
\caption{$m_V$ as a function of $m_{PS}^2$ for the symmetric
  representation at the values of $N$ shown. The solid lines are
  chiral fits to the data.}
\label{fig:2}
\end{center}
\end{figure}

\begin{figure}
\begin{center}
\includegraphics*[width=.6\textwidth]{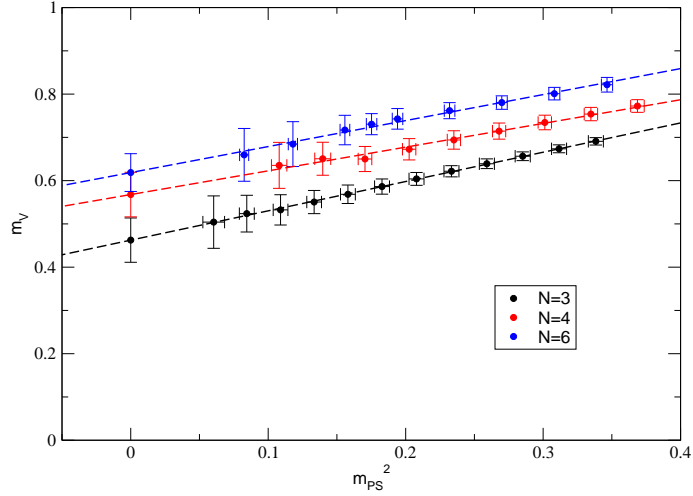}
\caption{$m_V$ as a function of $m_{PS}^2$ for the antisymmetric
  representation at the values of $N$ shown. The solid lines are
  chiral fits to the data.}
\label{fig:3}
\end{center}
\end{figure}

After extracting $m_{V}$ and $m_{PS}$, we have determined the chiral
limit value of $m_{V}$ (denoted as $m_V^{\chi}$) using the chiral ansatz
\begin{eqnarray}
\label{eq:fit}
m_V\left(m_{PS} \right) = m_V^{\chi} + B m_{PS}^2 \ .
\end{eqnarray}
Our results and the corresponding fits are displayed in
Figs.~\ref{fig:1},~\ref{fig:2}~and~\ref{fig:3} respectively for the
adjoint, symmetric and antisymmetric representations. We note that
there is a well-defined hierarchy for $m_V$ at fixed $m_{PS}$, with
the symmetric mass being the highest and the antisymmetric being the
lowest. At fixed representation, the vector mass in the antisymmetric has the largest
variation with $N$, while there is little or no variation with $N$ for the
adjoint vector masses.

\begin{figure}
\begin{center}
\includegraphics*[width=.6\textwidth]{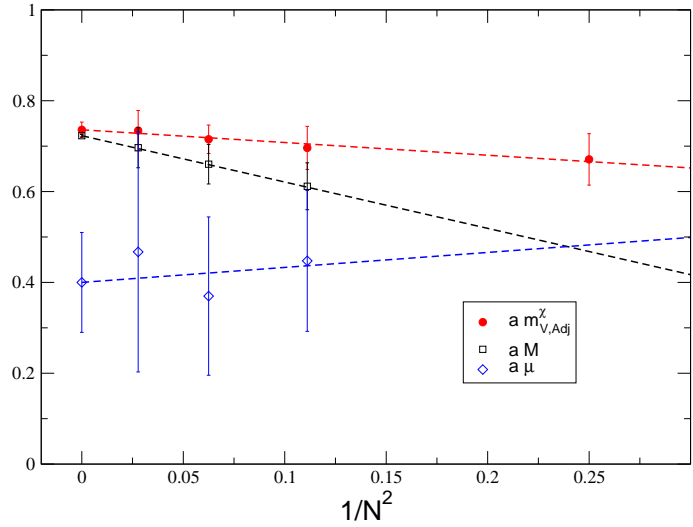}
\caption{
Large-$N$ extrapolation of the adjoint mass and of
the combinations $M$ and $\mu$,
all in the chiral limit.}
\label{fig:4}
\end{center}
\end{figure}

The large-$N$ extrapolation has been performed for the quantity
$m_{V,Adj}^\chi$ and for the quantities $M_V^\chi$ and $\mu_V^\chi$
(see Eqs.~\eqref{eq:auxmass}) using in all cases only the leading term
and the $1/N^2$ correction. The quality of the fits is displayed in
Fig.~\ref{fig:4}. The values of $m_{V,\Sym}^\chi$ and $m_{V,\AS}^\chi$ as
a function of $1/N$ are obtained inverting Eqs.~\eqref{eq:auxmass}.
Our results are:
\begin{gather}
m_{V,\Adj}^\chi = 0.736(17)  - \frac{0.28(11)}{N^2} \ , \\
m_{V,\Sym}^\chi = 0.723(27)  + \frac{0.40(11)}{N} - \frac{1.02(32)}{N^2} + \frac{0.3(1.2)}{N^3} \ , \\
m_{V,\AS}^\chi = 0.723(27)  - \frac{0.40(11)}{N} - \frac{1.02(32)}{N^2} - \frac{0.3(1.2)}{N^3} \ .
\end{gather}
The equality of the large-$N$ results for the three representations is a prediction of Orientifold
Planar Equivalence. We remark that through the separations of corrections
in even and odd powers of $1/N$ in the symmetric and antisymmetric
representation, we were able to get an handle on the
size of the corrections up to $O(1/N^3)$. Numerically, Orientifold
Planar Equivalence can be verified for any value of $m_{PS}$ in the
chiral region~\cite{Lucini:2010kj}.

\section{Conclusions}
Our work should be regarded as an illustration of a possible strategy
for a numerical proof of Orientifold Planar Equivalence. We have shown
that it is possible to combine the data in the symmetric and
antisymmetric representation for extracting precise values in the
large-$N$ limit from data at moderately small values of $N$.
The next step in this investigation is the simulation of the dynamical
theories.

\acknowledgments
We thank Luigi Del Debbio and Claudio Pica for the collaboration that lead to the development of the HiRep Monte Carlo code used in this work. Numerical simulations have been performed on a 120 core Beowulf cluster partially funded by the Royal Society and STFC and on the IBM Blue/C system at Swansea University. The work of B.L. is supported by the Royal Society through the University Research Fellowship scheme and by STFC under contract ST/G000506/1. A.R. thanks the Deutsche Forschungsgemeinschaft for financial support. A.P. was supported by the European Community - Research Infrastructure Action under the FP7 ``Capacities'' Specific Programme, project ``HadronPhysics2''.

\end{document}